\documentclass{ws-ijmpa}
\usepackage[super,compress]{cite}
\usepackage{graphics}
\usepackage{graphicx}
\usepackage{dcolumn}
\usepackage{bm}
\usepackage{mathrsfs}
\usepackage{pstricks}
\usepackage{color}
\usepackage{slashed}
\usepackage{amsmath}
\usepackage{epsfig}
\usepackage{amsfonts}
\usepackage{amssymb}
\def\beq{\begin{equation}}
\def\eeq{\end{equation}}
\def\bea{\begin{eqnarray}}
\def\eea{\end{eqnarray}}
\def\nn{\nonumber}

\def\Tr{\textrm{Tr}}
\def\x{\mathbf{x}}

\def\k{\mathbf{k}}

\def\p{\mathbf{p}}

\begin{document}

\markboth{G. Menezes, B. F. Svaiter, N. F. Svaiter}
{One-loop effective action and the Riemann Zeros}

%
\catchline{}{}{}{}{}
%

\title{ONE-LOOP EFFECTIVE ACTION AND THE RIEMANN ZEROS}
\author{J. G. DUE\~NAS}
\address{Centro Brasileiro de Pesquisas F\'{\i}sicas, Rio de Janeiro, RJ 22290-180, Brazil\\
jgduenas@cbpf.br}
\author{N. F. SVAITER}
\address{Centro Brasileiro de Pesquisas F\'{\i}sicas, Rio de Janeiro, RJ 22290-180, Brazil\\
nfuxsvai@cbpf.br}
\author{G. MENEZES}
\address{Departamento de F\'isica, Universidade Federal Rural do Rio de Janeiro, Serop\'edica, RJ 23890-000, Brazil\\
gabrielmenezes@ufrrj.br}

\maketitle

\begin{abstract}
We present a remarkable connection between the asymptotic behavior of the Riemann zeros and one-loop effective action in Euclidean scalar field theory. We show that in a two-dimensional space, the asymptotic behavior of the Fourier transform of two-point correlation functions fits the asymptotic distribution of the non-trivial zeros of the Riemann zeta function. We work out an explicit example, namely the non-linear sigma model in the leading order in $1/N$ expansion.
\end{abstract}

\ccode{PACS numbers: 02.10 De, 11.10.-z}

\section{Introduction}

Recently there have been investigations whether a large class of systems with countable infinite number of degrees of freedom
described by self-adjoint operators can be used to give a spectral interpretation for the Riemann zeros.
The main result we emphasize is that in the formalism of quantum field theory there is no obstruction for the
existence of the above mentioned elusive operators.

In this paper we delve a different point of view aiming to shed some light on this problem. We consider a large class of
two-dimensional field-theory models in Euclidean space. We show that the asymptotic behavior of the Fourier
transform of the two-point correlation function in such models fits the asymptotic distribution of the non-trivial zeros
of the Riemann zeta function. As an example we show that the non-linear sigma model in the leading order in $1/N$ expansion
presents such a behavior. Although there is in the literature intense activity connecting the Riemann zeros, classical and
quantum chaos, random matrices and disordered systems, as far as we know the result presented here is discussed for the first time.

The Riemann zeta function $\zeta(s)$ defined by analytic continuation of a Dirichlet series has a simple pole with
residue $1$ at $s=1$, trivial zeros at $s=-2n$, $n=1,2,...$ and complex zeros $\rho=\beta+i\gamma$ for $0<\beta<1$~\cite{riem,ingham}. The
Riemann hypothesis is the conjecture that $\beta=1/2$. Hilbert and P\'olya suggested that a possible way to prove the Riemann hypothesis would
be to interpret the non-trivial zeros spectrally. In this scenario one must find a Hermitian operator whose eigenvalues are the imaginary part
of the non-trivial zeros of the zeta function.

In a series of papers the Hilbert-P\'olya conjecture (as the above approach became known) has been studied in the context of quantum field theory.
Using the construction of the so-called super-zeta or secondary zeta function built over the Riemann zeros~
\cite{superzeta}, it was shown that it is possible to extend the Hilbert-P\'olya
conjecture to systems with countably infinite number of degrees of freedom~\cite{gn2}.
Based on the above discussed ideas, Due\~nas and Svaiter investigated the consequences of the existence of such
an operator in a quantum-field-theory framework. These authors studied a massless scalar field in a $(d+1)$-dimensional space-time,
where one of the coordinates lies in an finite interval $[0,a]$ in one direction, assuming that the the non-trivial zeros of the
zeta function appears as a subset of the spectrum of the vacuum modes. The renormalized zero-point energy of this system was
presented using an analytic regularization procedure~\cite{ds}.

Another route to exploit connections between number theory and quantum theory is known as arithmetic quantum field
theory~\cite{stn}. In these theories, the partition function of hypothetical systems are related to the Riemann zeta function or to
more general Dirichlet series. One example of such system is known as the bosonic Riemann gas~\cite{bakas,spec3}.
Recently, it was presented a generalization of such a model by assuming an ensemble of Hamiltonians of
arithmetic bosonic gas with some probability distribution~\cite{duenas}.
The singularity structure for the average free energy density of this arithmetic gas in the complex $\beta$ plane was discussed.

The connection between Riemann zeros, chaotic systems, disordered systems and random matrices has been widely discussed in
the literature \cite{for1,for2}. It was shown that the density of the Riemann zeros can be written as a trace formula where the role of periodic orbits is played by the sequence of the prime numbers. For a recent discussion see for example the Ref.~\refcite{bou}.
Other important result is the random matrix conjecture which claims that the energy levels of classically
chaotic systems are distributed as eigenvalues of random matrix ensembles~\cite{prl,bohigas,berry}.  The connection between Riemann zeros and random
matrices is given by the Montgomery-Odlyzko law which states that the distribution
of spacing between nontrivial zeros of the Riemann zeta function $\zeta(s)$ is statistically identical to the distribution of
eigenvalue spacings in a Gaussian unitary ensemble \cite{mo,mo2}. This is equivalent to the result that
the statistical asymptotic distribution of the Riemann zeros on the critical line coincides with the statistical distribution
of the eigenvalues of very large size random unitary matrices.

We briefly mention that non-linear sigma models can be brought to the context of random matrices and systems with disorder.
This is achieved by considering the fact that a gas of fermions in the presence of a random potential can be described by an
effective low-energy field theory, which is precisely a non-linear sigma model \cite{efetov1,efetov2}. It can be shown that
random matrix theory appears in an appropriate limit of such a model. The non-linear sigma model also emerges if we consider
the problem of wave propagation in a medium with a correlated spatially varying index of refraction \cite{john}. Since disorder
in wave problems leads us to chaotic systems in the limit of short wavelengths, one has an example in which a link between
the Riemann zeta zeros and the non-linear sigma model must be investigated.

The organization of the present paper is the following. We briefly review the connection between Riemann zeros and quantum
mechanics in Section II. We discuss the impossibility of extending such a scenario in a straightforward way to systems
with infinite number of degrees of freedom. In Section III we investigate if scalar field-theory models where
the loopwise expansion is available can provide some new insights in number-theory questions. In Section IV we show how the non-linear sigma model,
defined in a two-dimensional Euclidean space in a $1/N$ expansion is connected with the asymptotic distribution of the non-trivial zeros of the
Riemann zeta function. Conclusions are given in Section V. In this paper we use $k_{B}=c=\hbar=1$.

\section{Riemann zeros in quantum mechanics}

Berry and Keating conjectured that the Hamiltonian
\beq
H = \frac{1}{2}\left(q p + p q\right)
\label{be}
\eeq
has important connections with the non-trivial Riemann zeros~\cite{berry2,berry3}. Possible modifications and extensions to this interesting
scenario were proposed~\cite{sierra1,sierra2,sre}. In the context of quantum mechanics, such remarkable results are in some sense well
understood. However, it turns out that a simple generalization to quantum field theory leads one to rather suspicious consequences. The
argument goes as follows. A straightforward extension of the supersymmetric Wess-Zumino model~\cite{wz} to spaces of dimension $d > 4$ leads
one to equations of motion of higher order such as
\beq
\left(\Box^{\kappa/2} - m^{\kappa}\right)\varphi = 0,
\label{susy}
\eeq
where $\kappa = 2^{d/2-1}$ and $\varphi$ is a scalar quantum field. As discussed in Ref.~\refcite{bol}, these equations for any
$d > 4$ have a tachyonic component. Being more specific, for the case of six dimensions, the above equation can be derived
from the second-order Lagrangian
\beq
{\cal L} = \frac{1}{2}\left( \Box \varphi \Box \varphi - m^4\varphi^2\right),
\eeq
whose associated Hamiltonian has a tachyonic contribution. Such a term can be understood as representing a superposition
of systems having for each momentum degree of freedom $\k$ the following Hamiltonian
\beq
H_{\k} = \frac{1}{2}\left(q_{\k} p_{\k} + p_{\k} q_{\k}\right).
\label{bol}
\eeq
For more details see Ref.~\refcite{bol}. It is easy to see that equation~(\ref{bol}) is the straightforward generalization
of the Hamiltonian~(\ref{be}) to the case of infinite number of degrees of freedom. This leads us to a model for
tachyon quantization. As well known, such a model is plagued with many difficulties such as imaginary mass, lack of
unitarity, causality issues, etc. Therefore, as long as quantum field theory is concerned, models based on Eq.~(\ref{bol}) should be discarded on physical grounds.

A natural question now arises. How can one approach the Hilbert-P\'olya conjecture in the context of quantum field
theory? In the next Sections we show that there is a connection between the asymptotic distribution of the Riemann zeros and two-point
Schwinger functions of one-loop effective action in quantum field theory.

\section{The one-loop effective action}

In this Section we digress on the fundamental object through which correlation functions in many models
in two-dimensional Euclidean scalar field theory relate to the asymptotic distribution of Riemann zeros. We consider massive
interacting fields $\phi$ with (bare) mass $m$. Here the basic quantity we are interested in is
\beq
{\cal S}_{\textrm{eff}}[M]= \Tr\ln\left[-\partial^2 + M^2(m, x)\right],
\label{eff4}
\eeq
where $M(m, x)$ is some scalar function with dimensions of mass squared. Whenever a formal perturbative expansion in
powers of a suitable parameter is available, the usual Feynman diagrams of field theory can be displayed. In particular,~(\ref{eff4})
generates one-loop diagrams, in particular bubble diagrams. The quantity~(\ref{eff4}) naturally appears in the semi-classical or loopwise
expansion of field theory. As well known, this method amounts to introduce a formal expansion of the functional integral in powers of $\hbar$:
\beq
Z[J] = \int[d\phi]\exp{-\frac{1}{\hbar}\left\{S[\phi] - \int d^2x\,J(x)\phi(x)\right\}}
\eeq
where $[d\phi]$ is a formal product of Lebesgue measures at every point of $\mathbb{R}^{2}$, $J(x)$ is an external source and $S[\phi]$ is the action given by
\beq
S[\phi] = \int d^2x\,\biggl[\frac{1}{2}\Bigl(\partial_{\mu}\phi\partial_{\mu}\phi + m^2\phi^2\Bigr) + V[\phi]\biggr].
\eeq
Here we regard $V[\phi]$ as a simple polynomial in $\phi$. In addition, notice that one must return with the $\hbar$ in
front of the action in the functional integral. We are not going to present the standard derivation since it can be
found in many textbooks, see for instance Ref.~\refcite{zinn}. The important point is that, at order $\hbar$ one obtains
the following expression for the generating functional $\Gamma$ of proper vertices:
\beq
\Gamma[\varphi] = S[\varphi] + \frac{1}{2}\left\{{\cal S}_{\textrm{eff}}[M] - \Tr\ln\left[-\partial^2 + m^{2}\right]\right\},
\eeq
where $M^2 = m^2 + V''[\varphi]$, $\varphi$ being the classical field, solution of the classical field equations $\delta S[\varphi]/\delta\varphi = 0$.
The $n$-th functional derivative of $\Gamma$ with respect to $\varphi$  for $n \geq 3$ yields the one-particle irreducible correlation functions of the theory.

Let us present one situation where the basic quantity~(\ref{eff4}) shows up. Consider a simple model of two Euclidean
self-interacting scalar fields $\phi_1(x)$ and $\phi_2(x)$ with masses $m_1$ and $m_2$, respectively. Assume that
$m_2 \gg m_1$. After the construction of an effective field theory and further imposition of the infinite mass limit for
the heavy field, there is a decoupling between the light and heavy modes as stated by the Appelquist-Carrazzone
theorem~\cite{app}. This is essentially the model studied in Ref.~\refcite{nami} in two dimensions. Following such a
reference, we consider the theory described by the following action with two real scalar fields
\bea
S[\phi_1, \phi_2] &=& \int d^2x\,\biggl[\frac{1}{2}\Bigl(\partial_{\mu}\phi_1\partial_{\mu}\phi_1 + m_1^2\phi_1^2
+\, \partial_{\mu}\phi_2\partial_{\mu}\phi_2 + m_2^2\phi^2\Bigr)
\nn\\
&&+\, V[\phi_1]+ \frac{\lambda}{2}(\phi_1\phi_2)^2\biggr].
\eea
The precise form of $V[\phi_1]$ is not important for the construction of the effective theory, although it is essential in order to
consistently implement the decoupling theorem~\cite{nami}. The partition function is given by
\beq
Z[j_1, j_2] = N\int[d\phi_1][d\phi_2]\,\exp{\left\{-S[\phi_1, \phi_2]\right\}}
\exp{\left\{-\int d^2x\,(j_1\phi_1 + j_2\phi_2)\right\}},
\eeq
where $j_1(x), j_2(x)$ are external sources and $N$ is a normalization. The above action can be rewritten as
\beq
S[\phi_1, \phi_2] = S_1[\phi_1] + S_2[\phi_1, \phi_2],
\eeq
where $S_1[\phi_1]$ is the $\phi_2$-independent part of the total action. In order to obtain the Wilsonian effective action for
the light modes one must assume that $m_2 \gg m_1$. In this way we integrate out the heavy field $\phi_2$ in the
functional integral and define the effective action of the light modes $\Gamma_{\textrm{eff}}[\phi_1]$ by
\beq
e^{-\Gamma_{\textrm{eff}}[\phi_1]} = e^{-S_1[\phi_1]}\int[d\phi_2]\,e^{-S_2[\phi_1, \phi_2]}
\eeq
By means of a simple Gaussian integration, we obtain the following expression for the Wilsonian effective action of the light field $\phi_1(x)$
\beq
\Gamma_{\textrm{eff}}[\phi_1] = S_1[\phi_1] + \frac{1}{2}{\cal S}_{\textrm{eff}}[M],
\eeq
where $M^2 = m_2^2 + \lambda\phi_1^2$. So we see that ${\cal S}_{\textrm{eff}}$ here appears as the one-loop contribution to the
Wilsonian effective action for the light fields.

Another example in which Eq.~(\ref{eff4}) appears is provided by an $O(N)$ symmetry Euclidean field theory with a
$N$-component scalar field $\Phi(x)$. In a two-dimensional Euclidean space, the action is given by
\beq
S[\Phi]=\int d^{2}x \left\{\frac{1}{2}\partial_\mu\Phi^{T}\partial_\mu\Phi +N\,U\biggl[\frac{\Phi^2}{N}\biggr]\right\},
\eeq
where $U$ is a general polynomial in $\Phi^2$ and
$$\Phi^2=\sum_{i=1}^{N}\varphi_{i}^{2}.$$
The correlation functions of the model are generated by the partition function
\beq
Z[h]=\int\,\prod_{x}d\Phi(x)\,\exp\biggl[-S[\Phi]+ \int\,d^{2}x\,h^{T}(x)\Phi(x)\biggr],
\eeq
where the last term in the argument of the exponential is the contribution for an auxiliary $N$-component external field $h(x)$.
Since we are interested in finite theories, a cut-off consistent with the symmetries of the model is implicit. Introduce the field $\rho(x)$ defined by
\beq
\rho(x)=\frac{\Phi^2}{N}
\eeq
and the field $\lambda(x)$ which acts as a Lagrange multiplier, i.e.
\beq
1=\frac{N}{4\pi\,i}\int\,d\rho\,d\lambda\,\exp\bigg[\frac{\lambda}{2}\left(\Phi^{2}-N\rho\right)\biggr],
\eeq
where the $\lambda$ integration runs parallel to the imaginary axis.  Using the fields  $\lambda(x)$ and $\rho(x)$, and
defining the respective functional measures $[\rho(x)]$ and $[\lambda(x)]$ we can write the partition function $Z[0]$ as
\beq
Z[0]= \int[d\Phi]\,[d\rho]\,[d\lambda]\exp{\{-S[\Phi, \rho, \lambda]\}},
\eeq
where $S[\Phi, \rho, \lambda]$ is given by
\beq
S[\Phi,\rho,\lambda] = \int d^{2}x\,\biggl\{\frac{1}{2}\bigl(\partial_{\mu}\Phi\bigr)^{2}+N\,U\bigl[\rho(x)\bigr]
+\frac{1}{2}\lambda(x)\bigl[\Phi^{2}(x)-N\rho(x)\bigr]\biggr\}.
\eeq
Let us write the $N$-component field $\Phi(x)$ as
$$\Phi^{T}=(\pi^{1},\pi^{2},...,\pi^{N-1},\sigma).$$
Integrating out the $N-1$ components $\pi(x)$ and introducing a one-component source $h(x)$ we have
\beq
Z[h] = \int[d\sigma][d\rho][d\lambda]\exp\biggl[-S_{N}(\sigma,\rho,\lambda)
+\int\,d^{2}x\,h(x)\sigma(x)\biggr],
\eeq
where
\bea
S_{N}(\sigma,\rho,\lambda)&=&\int d^{2}x\,\biggl\{\frac{1}{2}\bigl(\partial_{\mu}\sigma\bigr)^{2}+N\,U\bigl[\rho(x)\bigr]
+\frac{1}{2}\lambda(x)\bigl[\sigma^{2}(x)-N\rho(x)\bigr]\biggr\}
\nn\\
&+& \frac{1}{2}(N-1) {\cal S}_{\textrm{eff}}[M],
\eea
where now $M^2(x) = \lambda(x)$. In the large $N$ expansion we can obtain the saddle point equations. We are looking
for a uniform saddle-point: $\sigma(x)=\sigma$, $\rho(x)=\rho$ and finally $\lambda(x)=m^{2}$. We are particularly
interested in study fluctuations around the uniform saddle-point solution. One can show that the asymptotic behavior
of the Fourier transform of the two-point correlation function of the linear sigma model in the leading order in $1/N$
expansion in a two-dimensional Euclidean space fits the asymptotic distribution of the Riemann zeros. In the next section
we will present explicit calculations that prove our statement.

\section{The non-linear Sigma model and the Riemann zeros}

The sigma model was first introduced in Ref.~\refcite{gell} as an example realizing chiral symmetry and partial conservation
of the axial current. The model is constructed with a fermionic isodoublet field, a triplet of pseudoscalar fields and also a scalar field.
The sigma model without a fermionic isodoublet field with only two scalar fields $\sigma$ and $\pi$ has widely discussed
in the literature, as an example of spontaneous symmetry breaking.
This model can be generalized using a $N$-component field, in such a way that the Lagrangian density is
invariant under the orthogonal group in $N$-dimension, $O(N)$ \cite{dolan}.
The local action of the model in a four-dimensional Euclidean space is
\beq
S[\Phi]=\frac{1}{2}\int d^{4}x \left[\partial_\mu\Phi^{T}\partial_\mu\Phi +V(\Phi^2)\right]
\eeq
%
with $\Phi^{T}=(\pi^{1},\pi^{2},...,\pi^{N-1},\sigma)$ and $\Phi^{2}=\Phi^{T}\Phi$. The potential $V(\Phi^{2})$ is
chosen to produce a minimum whenever $\Phi^{2}=v^{2}>0$.
The system exhibits spontaneous symmetry breaking. There is a non-trivial
subgroup which leave the vacuum invariant, the $O(N-1)$. For each generator of the $O(N)$ group which does not leave the
vacuum invariant, there corresponds a massless Goldstone boson. An alternative description to produce spontaneous symmetry
breaking is given by the non-linear sigma model ~\cite{col}. The non-linear sigma model is defined by the Lagrangian density
\beq
\mathcal{L}=\frac{1}{2}\partial_\mu\Phi^{T}\partial_\mu\Phi
\eeq
with the constraint $\Phi^{T}\Phi=1$. The generating functional of the Schwinger functions can be written as
\beq
Z=\int[d\Phi]\delta[\Phi^{T}\Phi-1]\exp\left(\int d^{4}x\,\mathcal{L} + \textrm{source terms}\right).
\eeq
The quantization of non-linear sigma models was initiated in Refs.~\refcite{hon,ger,fad}. Although this model is not renormalizable
in a four dimensional Euclidean space, it make sense as a low-energy effective theory. The two dimensional case is quite different.
In this situation the model is perturbatively renormalizable. One can show that the beta function of the model is negative, therefore
the theory is also asymptotically free for $N>2$. The renormalizability of the two-dimensional non-linear sigma model was proved in Refs.~\refcite{bre,bec}.

Now we present some evidence that the large $N$ expansion of the two-dimensional $O(N)$ non-linear sigma model described by the Euclidean action
\beq
S[\Phi] = \frac{1}{2g^{2}_{0}}\int d^{2}x\, \partial_{\mu}\Phi^{T}\partial_{\mu}\Phi,\,\,\,\,\,\Phi^{T}\Phi = 1
\label{ac}
\eeq
might have a deep connection with the asymptotic distribution of the Riemann zeros. For a complete review of the large $N$
expansion in quantum field theory, see the Ref.~\refcite{mm}. In equation~(\ref{ac}) $g_{0}$ is a coupling constant and the
field $\Phi$ is a $N$-component vector. We are following the discussion presented in Refs.~\refcite{polyakov,tsv}. The partition
function which describes such modes in this model reads
\bea
&Z& = \int[d\Phi]\delta[\Phi^{T}\Phi-1]\exp{\left[-\frac{1}{2g_{0}^{2}}\int d^{2}x\,\, \partial_{\mu}\Phi^{T}\partial_{\mu}\Phi\right]}
\nn\\
&=& \int_{_{C-i\infty}}^{^{C+i\infty}} [d\lambda(x)]\int[d\Phi]\,\exp{\left\{-\frac{1}{2g_{0}^{2}}\int d^{2}x\,
\left[\partial_{\mu}\Phi^{T}\partial_{\mu}\Phi + \lambda(x)\left(\Phi^{T}\Phi - 1\right)\right]\right\}}
\label{imp}
\eea
where a Lagrange multiplier $\lambda(x)$ was introduced to substitute the functional Dirac delta function. The integral over the the
field $\Phi$ can be performed, since all of them are Gaussian integrals. We can write the partition function as
\beq
Z =\int_{_{C-i\infty}}^{^{C+i\infty}} [d\lambda(x)]\,e^{-S[\lambda]},
\label{imp2}
\eeq
where the action $S[\lambda]$ is given by
\beq
S[\lambda] = -\frac{1}{2g_{0}^{2}}\int d^{2}x \lambda(x) + \frac{N}{2}\ln\det\left[-\partial^2 + \lambda(x)\right].
\label{eff}
\eeq
As discussed in Ref.~\refcite{polyakov}, the expansion above is formal since we have infrared
and also ultraviolet divergences. To deal with the ultraviolet divergences we can introduce a cut-off or define the
theory in a lattice~\cite{kupianen}. Another approach is to use analytic regularization procedure. For an instructive example where
an analytic regularization procedure is used to evaluate the Fredholm determinant, see the Ref.~\refcite{nami}.
To cure the infrared divergences one can consider that the theory is defined in a finite volume.
Introducing the Green function $G(x,x;\lambda)$ defined as
\beq
G(x,x')=\langle\, x|(-\partial^{2}+\lambda)^{-1}|\,x'\,\rangle,
\eeq
and computing the variation of Eq. (\ref{imp2}) with respect to $\lambda$ we get
\beq
\frac{1}{2g_{0}^{2}}=\frac{N}{2}G(x,x;\lambda).
\eeq
In the saddle point approximation, using a Fourier representation for $G(x,x';\lambda)$ we can solve the equation above.
The cases $d=2$ and $d>2$ must be studied separately. We are interested only in the case $d=2$, where
continuous symmetries cannot be broken. The $O(N)$ symmetry is unbroken, since the expectation value for the components of
$\Phi$ are zero, but the field $\lambda(x)$ has a nonzero vacuum expectation value, i.e. $\langle\,\lambda(x)\,\rangle=m^{2}$.
We identify this quantity as a squared mass. The excitations associated with the components for the $\Phi$ field are massive particles. We are interested in studying fluctuations around this solution. With the identification
\beq
\lambda(x)=m^2+i\alpha(x),
\eeq
we get the effective action $S_{\textrm{eff}}$ 
\beq
S_{\textrm{eff}}[\alpha] = \frac{N}{2}\Tr\ln\left[-\partial^2 + m^{2}+i\alpha(x)\right].
\label{eff2}
\eeq
Compare with equation~(\ref{eff4}). Introduce the Fourier expansion
\beq
\alpha(x) = \int\frac{d^2p}{(2\pi)^2}\,\alpha(\p)\,e^{i\p\x}
\eeq
and let us consider the effective action for the modes $\p \neq 0$. Using a functional Taylor series one gets
\bea
S_{\textrm{eff}}[\alpha] &=& \frac{N}{2}\Tr\ln\left[\Delta^{-1}\right]
+\frac{1}{2}\int\,\frac{d^2p}{(2\pi)^2} u(-p)\Pi(p)u(p)
\nn\\
&+&\sum_{n=3}^{\infty}\frac{(-1)^{n+1}}{n}\left(\frac{N}{2}\right)^{-n/2 + 1}\Tr\left[\left(iu(x)\Delta\right)^{n}\right],
\label{eff3}
\eea
where we have defined the field
\beq
u(x)=\left(\frac{N}{2}\right)^{\frac{1}{2}}\alpha(x)
\eeq
and the propagator
\beq
\Delta(x,x')=\langle\, x|(-\partial^{2}+ m^2)^{-1}|\,x'\,\rangle.
\eeq
Also, in equation~(\ref{eff3}) we have the quantity
\beq
\Pi(p) = \int\frac{d^2q}{(2\pi)^2}\Delta(p/2 + q)\Delta(p/2 - q),
\eeq
where
\beq
\Delta(p) = \frac{1}{p^2 + m^2}.
\eeq
The above series expansion furnishes a perturbative expansion for the $u$-field in the parameter $1/\sqrt{N}$. The zeroth-order
effective action for the fluctuations is given by the quadratic terms in the mentioned expansion. Hence in the leading order, the
Fourier transform of the two-point correlation function for the $u$-fields is given by
\beq
\langle u(-p)u(p) \rangle = \left[\Pi(p)\right]^{-1}.
\label{corr-func}
\eeq
Let us present an expression for this correlation function at large $|p| \gg m$. Consider the integral
\beq
I(\alpha,\beta,d,p) = \int\frac{d^d q}{(2\pi)^d}\frac{1}{\left[(q- p)^2 + m^2\right]^{\alpha}\left[q^2 + m^2\right]^{\beta}}.
\eeq
It is easy to see that $I(1,1,2,p) = \Pi(p)$. After standard manipulations we get
\bea
I(\alpha,\beta,d,p) &=& \frac{1}{(4\pi)^{d/2}(p^2)^{\alpha + \beta - d/2}}\frac{\Gamma(\alpha + \beta - d/2)}{\Gamma(\alpha)\Gamma(\beta)}
\nn\\
&&\times\int_{0}^{1}dx\,x^{\alpha-1} (1-x)^{\beta - 1}\left[x(1-x) + \frac{m^2}{p^2}\right]^{d/2-\alpha-\beta}
\eea
where $\Gamma(x)$ is the usual gamma function. Therefore, after an elementary integration we get
\bea
\Pi(p) &=& \frac{1}{2\pi f(m/p) p^2}\ln\left[\frac{1 + f(m/p)}{-1 + f(m/p)}\right],
\eea
where
\beq
f(x) = \left(1 + 4 x^2\right)^{1/2}.
\eeq
Since we are interesting in the asymptotic result, let us assume that $|p| \gg m$. Under this assumption we get
\beq
\langle u(-p)u(p) \rangle|_{|p| \gg m} \approx \frac{2\pi t}{\ln(t/m^2)},
\label{ns}
\eeq
where $t = t(p) = p^2 = p_0^2 + p_1^2$ is the equation of a circular paraboloid. We stress that the theory is finite after a renormalization
procedure. Note that this result obtained in Eq.~(\ref{ns}) has a remarkable resemblance with the asymptotic distribution of Riemann zeros.
Remember that the number $N(T)$ of Riemann zeros $\sigma + i\gamma$, with $0 < \gamma < T$ is asymptotically given by the Riemann-von Mangoldt formula:
\beq
N(T) = \frac{T}{2\pi}\ln\left(\frac{T}{2\pi}\right) - \frac{T}{2\pi} + O(\ln T),
\eeq
as $T\to\infty$. The density $D(T) = N(T+1) - N(T)$ of zeros at height $T$ is given by
\beq
D(T) = \frac{1}{2\pi}\left[\ln\left(\frac{T}{2\pi}\right)\right] + O\left( 1/T \right).
\eeq
An important consequence of such a result is that the imaginary parts of consecutive zeta zeros in the
upper half-plane $0 < \gamma_1 \leq \gamma_2 \leq \gamma_3 \leq \cdots$ satisfy $\gamma_n \sim n/D(n)$ as $n\to\infty$, i.e.
\beq
\gamma_n \sim \frac{2\pi n}{\ln\left(n/2\pi\right)}.
\label{zz}
\eeq
The similarity between the Eq.~(\ref{ns}) and Eq.~(\ref{zz}) is manifest. We have established the asymptotic equivalence between the distribution of the non-trivial zeros of the Riemann zeta function and the Fourier transform of the two-point correlation function of the non-linear sigma model in the leading order in a large $N$ expansion defined in a two-dimensional Euclidean space. Here the
identification $n \leftrightarrow t$ leads one to the conclusion that $n$ asymptotically obeys a circular paraboloid equation in the momentum space of a field theory. The results of this paper show us that there is a deep connection between number theory and quantum field theory, that deserves a further investigation.

\section{Conclusions}

The non-linear sigma model in a two-dimensional Euclidean space was used as a toy model for the study of asymptotic freedom
and dynamical mass generation \cite{poli}. In this paper we established a precise connection between the asymptotic distribution of the Riemann zeros and the asymptotic behavior of the Fourier transform of the two-point correlation function of the non-linear sigma model in the leading order in $1/N$ expansion in a two-dimensional Euclidean space.

As far as we know, although the literature emphasizes connections between the Riemann zeros, chaos, random matrices and disordered systems,
the result presented above is original and sound. Observe that in the semiclassical periodic orbit theory, the oscillating density of states
is related though a trace formula to classical periodic orbits \cite{gutzwiller1,gutzwiller2}. It is possible to obtain trace formulas
for the Riemann zeta function. Assuming the Riemannn hypothesis there is a remarkable parallel between the trace formulas of chaotic
systems and trace formulas for the Riemann zeta function. This analysis is of order $\hbar$ as the large $N$ expansion in the scalar field theory.

Of particular interest is to find other models where the behavior is similar to the linear and non-linear sigma model. One model that
calls for attention is the $CP(N-1)$ model, which can be solved in the large $N$ limit \cite{CP}. Also, since two-dimensional quantum
chromodynamics (QCD) reduces to the non-linear sigma model in some limit \cite{falomir}, the result of this paper indicates that a
connection between the Riemann zeros and QCD must be investigated. On the other hand a pivotal matter that confronts us is the
nature of $t(p)$. The Riemann zeros form a discrete set of points whereas for the quantum field theory considered here $t$ is
a continuous variable. This difficulty is already encountered within the original Berry-Keating approach~\cite{berry2}. At the present
moment we do not know how to circumvent this obstacle. These important subjects are reserved for future investigations.

\section*{Acknowlegements}

We thank J. P. Keating, Benar Svaiter, Jos\'e Ram\'on Mar\'{i}, Sebasti\~ao Alves Dias, T. Micklitz and Jorge Stephany Ruiz for useful discussions.
This paper was supported by Conselho Nacional de Desenvolvimento Cientifico e Tecnol{\'o}gico do Brazil (CNPq).

\end{document}